\documentclass[]{spie}  %>>> use for US letter paper
\usepackage[colorlinks=true, allcolors=blue]{hyperref}
\usepackage{amsmath,amsfonts,amssymb}
\usepackage{graphicx}
\usepackage{subcaption}
\usepackage{color}

\usepackage[utf8]{inputenc}
\usepackage[T2A,T1]{fontenc}
\usepackage[french,russian,polish,english]{babel}
\usepackage{CJKutf8} %\usepackage{CJK}
\author{Brendan Jou}
\author{Shih-Fu Chang}
\affil{Department of Electrical Engineering, Columbia University, New York, NY 10027}

\authorinfo{Primary correspondence author: B.~Jou.\\E-mails: \{bjou,sfchang\}@ee.columbia.edu}

\title{Going Deeper for Multilingual Visual Sentiment Detection}

\begin{document}

\maketitle


\begin{abstract}
This technical report details several improvements to the visual concept detector banks built on images from the Multilingual Visual Sentiment Ontology (MVSO) \cite{jou_2015}.
The detector banks are trained to detect a total of 9,918 sentiment-biased visual concepts from six major languages: English, Spanish, Italian, French, German and Chinese.
In the original MVSO release \cite{jou_2015}, adjective-noun pair (ANP) detectors were trained for the six languages using an AlexNet-styled architecture \cite{krizhevsky_2012} by fine-tuning from DeepSentiBank \cite{chen_2014_dsb}.
Here, through a more extensive set of experiments, parameter tuning, and training runs, we detail and release higher accuracy models for detecting ANPs across six languages from the same image pool and setting as in the original release using a more modern architecture, GoogLeNet \cite{szegedy_2015}, providing comparable or better performance with reduced network parameter cost.

In addition, since the image pool in MVSO \cite{jou_2015} can be corrupted by user noise from social interactions, we partitioned out a sub-corpus of MVSO images based on tag-restricted queries for higher fidelity labels.
We show that as a result of these higher fidelity labels, higher performing AlexNet-styled \cite{krizhevsky_2012} ANP detectors can be trained using the tag-restricted image subset as compared to the models in full corpus.
We release all these newly trained models for public research use along with the list of tag-restricted images from the MVSO dataset.
\end{abstract}

\section{INTRODUCTION}
\label{sec:intro}

Following the trend in many other fields, the advent of high-volume and weakly-supervised data is driving increased interest in \emph{large-scale} sentiment studies in Affective Computing \cite{picard_1997}.
However, directly studying affect, or emotion, in a dimensional representation (e.g.~valence-arousal-dominance) or even discrete semantics (e.g.~sad, happy, angry, etc.) tend to suffer from the problem of data sparsity since such specialized psychology terminology are less likely to be found in large volume from the Web.
Inspired by similar work in Computer Vision and Multimedia, several works have since proposed affective mid-level representations to bridge the \emph{affective gap} between low-level features and high-level affect states.
One such attractive mid-level semantic representation is the \emph{adjective-noun pair} (ANP) \cite{borth_2013_vso}, where `nouns' are used to provide visual grounding to a detectable object and `adjectives' are used to sentimentally bias the object.

In a recent work, we presented a large-scale multilingual visual sentiment ontology (MVSO) \cite{jou_2015} consisting of 15,630 such ANPs across 12 languages with an associated dataset of over 7.36 million images.
As part of this initial work, we developed a set of language-specific visual concept detector banks that could identify the presence of ANPs in six of the major languages in the MVSO.
These detector banks were shown to be useful in cross-lingual sentiment analysis \cite{jou_2015}, and recently, in diversifying image query expansion \cite{li_2016} and facial imagery understanding \cite{pappas_2016}.
Yet in the original development of the ANP detectors, many of the original training parameter settings were defaulted and left as-is for a proof-of-concept.
Even though using such default settings show that the multilingual ANP detectors can achieve reasonable detection performance already, they do not represent the most near-optimal networks that can be trained for the given architectures used.

In this report, we detail and release
(1) higher accuracy models for detecting adjective-noun pairs (ANPs) in six languages from the same image pool in the original Multilingual Visual Sentiment Ontology (MVSO) release \cite{jou_2015} and new detector banks fine-tuned using a different more modern network architecture with benchmark comparisons,
(2) a sub-corpus of MVSO based on tag-restricted queries for higher fidelity labels, and
(3) adjective-noun pair detectors based on this tag-restricted subset.
The model and data release can be found through the original MVSO website at \url{http://mvso.cs.columbia.edu}.

\section{BRIEF OVERVIEW OF MVSO}
\label{sec:mvso_overview}

The Multilingual Visual Sentiment Ontology (MVSO) \cite{jou_2015}, a substantial extension and improvement on the Visual Sentiment Ontology (VSO) \cite{borth_2013_vso}, consists of over 15.6K sentiment-biased mid-level visual concepts using the semantic construct of adjective-noun pairs (ANPs).
The ontology was constructed by first using seed emotion keywords from ``Plutchik's Wheel of Emotions'' \cite{plutchik_1980} where were translated into 12 different languages by native speakers.
For each language, these keywords were used to query the Flickr API\footnote{\url{https://www.flickr.com/services/api}} to retrieve a large corpus of images with related tags and other metadata.
From the tags and metadata, we mined popularly used adjectives and nouns, identified by language-dependent part-of-speech taggers, combinatorially formed adjective-noun pair (ANP) candidates, and then filtered them on a number of criteria including semantic correctness, sentiment strength, popular usage on Flickr, and coverage in Flickr uploader diversity.
After this candidate filtering, the MVSO consisted of 15,630 ANP concepts representative of 12 different languages (language codes in brackets): Arabic {\bf [ar]}, Chinese {\bf [zh]}, Dutch {\bf [nl]}, English {\bf [en]}, French {\bf [fr]}, German {\bf [de]}, Italian {\bf [it]}, Persian {\bf [fa]}, Polish {\bf [pl]}, Russian {\bf [ru]}, Spanish {\bf [es]} and Turkish {\bf [tr]}.
These ANP concepts were then used to query the Flickr API to retrieve associated images and metadata, and it was later found that our ANPs also have geo-reference coverage of over 235 countries \cite{jou_2016_senticart}.

\begin{figure}[h]
  \centering
  \includegraphics[width=5.3in]{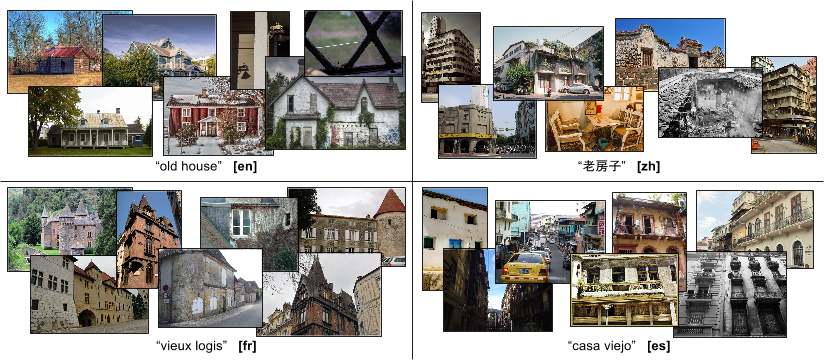}
  \caption{Example MVSO Images in Four Sampled Languages from a Multilingual ANP Cluster. The cluster, formed by approximate matching, portrays concepts related to the ``old'' or ``ancient'' nature of some ``houses''.}
  \label{fig:old_house}
\end{figure}

To develop an actual ontology with a semantic tree structure, we proposed two construction methods: \emph{exact matching} and \emph{approximate matching} \cite{jou_2015,pappas_2016}.
In `exact matching,' we use a pivot language where all ANPs are automatically translated into English and ANPs become naturally grouped by their translations.
This yields high cluster ``precision,'' but low ``recall'' since nuances like plurals and synonyms, and even related concepts are not grouped together.
In `approximate matching,' we extract word feature vectors from the ANPs and perform traditional distance-based clustering.
This trades of some cluster consistency for increased cluster coverage.
These clustering approaches have proven useful in uncovering some multilingual geographic social phenomena \cite{jou_2016_senticart} as well as in cross-lingual sentiment analysis and query expansion \cite{jou_2015,li_2016}.
Samples images from MVSO for an ANP cluster with concepts like `old house' are shown in Figure \ref{fig:old_house}.

The final MVSO image corpus consists of 7,368,364 images from which language-specific ANP detector banks were trained.
Six of the major languages were chosen to train these detectors, resulting in a coverage of 9,918 total ANPs across 6,994,319 images in English {\bf [en]}, Spanish {\bf [es]}, Italian {\bf [it]}, French {\bf [fr]}, German {\bf [de]} and Chinese {\bf [zh]}.
For more details on MVSO, we refer the interested reader to the related peer-review publications \cite{jou_2015,jou_2016_senticart,li_2016,pappas_2016}.

\section{DEEP VISUAL SENTIMENT CONCEPT DETECTOR BANKS}

In our original MVSO publication \cite{jou_2015}, we trained six detector banks, fine-tuning using an AlexNet-styled network architecture where there are five convolutional layers and three fully-connected layers \cite{krizhevsky_2012} with pooling and normalization layers swapped to create a variant called CaffeNet \cite{jia_2014}.
We initialized our networks weights from a CaffeNet fine-tuned network called DeepSentiBank \cite{chen_2014_dsb}, which was trained to detect 2,089 ANPs (in English) from the images of the VSO dataset \cite{borth_2013_vso} and improving on its original proposed SVM-based detector banks, SentiBank \cite{borth_2013_sb}.
In fine-tuning these six detector banks, we followed the traditional fine-tuning strategy of resizing the final fully-connected layer to have the same number of output units as the number of classes with random initialization (in this case, Gaussian), started optimization at a lower base learning rate (in this case, 0.001), and used a higher learning rate multiplier for the last fully-connected layer to allow for more aggressive updates on that layer compared to the other layers that were pretrained (in this case, $\times$2).
During optimization, we used a batch size of 256 and fixed the learning rate, gradually reducing it by a factor of 10 whenever we saw training loss plateau.
Every other setting in the network architecture, e.g.~dropout rates and output sizes of other fully-connected layers, and training routines, e.g.~optimization technique, was kept the same as when training the original ILSVRC \cite{imagenet} model\footnote{See \url{https://github.com/BVLC/caffe/tree/master/models/bvlc_reference_caffenet} for parameter details.}.
We note that in these MVSO detectors, the original ILSVRC12 mean image was used for mean subtraction.
All training was originally performed on a single NVIDIA GeForce GTX 980 GPU and was implemented with Caffe \cite{jia_2014}.
The performances in the original MVSO release \cite{jou_2015} are copied in Table \ref{tab:old_classifiers}.

To further improve the detection performance of ANPs, we experimented with a much deeper and more complex architecture called GoogLeNet, or Inception \cite{szegedy_2015}.
The original GoogLeNet design consists of mini-networks, also called Inception modules, that are composed of 1$\times$1\emph{C}, 1$\times$1\emph{C}-3$\times$3\emph{C}, 1$\times$1\emph{C}-5$\times$5\emph{C}, and 3$\times$3\emph{MP}-1$\times$1\emph{C} towers where \emph{C} corresponds to convolutions and \emph{MP} denotes max pooling, i.e.~so 1$\times$1\emph{C}-3$\times$3\emph{C} denotes one path in the modules that has a 1$\times$1 convolution followed by a 3$\times$3 convolution.
While AlexNet-styled architectures are only 8 layers deep, the GoogLeNet architecture is 22 layers deep while still using less parameters\footnote{Note that here we use the first iteration of GoogLeNet/Inception, sometimes referred to as Inception-v1, though there have been several proposed improvements since that use double 3$\times$3 layers or that integrate residual learning.}.
These fan-in-fan-out modules are stacked in an architecture referred to as network-in-network \cite{lin_2014}.
GoogLeNet additionally uses two auxiliary branches in the network to prevent gradients from vanishing during backpropagation \cite{szegedy_2015}.

For training, we preserved the same training setting as in the original MVSO release \cite{jou_2015} using the same train/test split of images.
Unlike before though we pre-train from a ILSVRC12 model since the architecture is significantly different from DeepSentiBank \cite{chen_2014_dsb} and no comparable Inception-like network was previously trained on VSO.
In addition, since the output space for languages like English and Spanish is large, in the auxiliary heads of Inception, we widened the second-to-last fully-connected layers which were originally 1,024 to 4,352 and 4,096, respectively, since they would otherwise become a representational bottleneck.
For English and Spanish, we found that using the second auxiliary head as the final output actually yielded higher performance due to this layer widening and so we truncate the network after this second auxiliary head for the final network.
For all other languages, the auxiliary heads were truncated and the usual penultimate layers were preserved in the final network.
We used SGD with a sigmoid decay learning rate decay with a base learning rate of 0.001, a batch size of 64 and a decay factor of 0.1 and unlike before use the true training data mean image during mean subtraction.

\section{EXPERIMENTS \& DISCUSSION}

All new experiments used a single NVIDIA GeForce GTX Titan X GPU and were implemented with Caffe \cite{jia_2014}.

\begin{table}[t]
  \centering
  \begin{tabular}{lccccccc}
    \textbf{Language} & \textbf{\#ANPs} & \textbf{\#params} & \textbf{\#train} & \textbf{\#test} & \textbf{time} & \textbf{top-1} & \textbf{top-5} \\ \hline
    English & 4,342 & 74.66 & 3,236,728 & 807,447 & 95 & 10.13\% & 21.06\% \\
    Spanish & 2,382 & 66.63 & 1,085,678 & 270,400 & 45 & 12.35\% & 25.36\% \\
    Italian & 1,561 & 63.26 &   602,424 & 149,901 & 30 & 17.01\% & 30.93\% \\
    French  & 1,115 & 61.44 &   462,522 & 115,112 & 26 & 17.66\% & 35.46\% \\
    German  &   275 & 57.99 &   108,744 &  27,048 & 12 & 30.11\% & 52.78\% \\
    Chinese &   243 & 57.86 &   102,740 &  25,575 & 15 & 27.07\% & 45.06\%
  \end{tabular}
  \caption{Classification accuracies of the ANP detector banks (\%) from the original MVSO release\cite{jou_2015}, where images were collected from a hybrid pool of tag-restricted and free text searches. This table essentially mirrors Table 4 in the original publication in ACM Multimedia 2015\cite{jou_2015} except now to two-decimal precision and with estimated walltimes (hours) instead of CPU/GPU process timing and number of network parameters (millions).}
  \label{tab:old_classifiers}
\end{table}

\subsection{Hybrid-pool Multilingual ANP Detectors}
\label{ssec:hybridpool}

In the original MVSO image mining \cite{jou_2015}, when we downloaded data, we first queried the Flickr API using our multilingual ANPs.
Generally, there are two types of ways to search for photos on Flickr: asking the Flickr API to search for the query in the (1) tags of a photo, or (2) anywhere in the photo text data, including photo title, description and tags.
We call these ``tag search'' and ``free-text search,'' respectively.
Tag search yields less but more precise results, and free-text search will give more but noisier data, i.e.~leading to weaker supervision.
The intuition is that if a user uses an ANP as a tag, then it is more likely to describe what is actually visually present in the image, while an ANP that occurs in, say the description, may be relevant, but not visually present.
In our data collection, we used an upper bound of 1,000 images per ANP where we first queried via tag search and if the upper bound had not been reached, we then pulled results from free-text search on Flickr.
As a result, the original MVSO image dataset constitutes a \emph{hybrid-pool} of Flickr images with respect to how the querying was performed.
The six ANP detector banks presented in the original release \cite{jou_2015} were based on this very hybrid-pool of images.
In Table \ref{tab:old_classifiers}, we show the top-$k$ accuracies of the original \cite{jou_2015} CaffeNet-structured \cite{jia_2014}, DeepSentiBank-fine-tuned \cite{chen_2014_dsb} detector banks using this hybrid-pool of images.
Likewise, in Table \ref{tab:new_classifiers}, we show the top-$k$ performances of the new Inception-styled architectures for the same six languages on the same hybrid-pool of images.
Top-$k$ accuracy refers to the percentage of classifications for which the true class is in the top $k$ predicted ranks.

\begin{table}[h]
  \centering
  \begin{tabular}{lccccccc}
    \textbf{Language} & \textbf{\#ANPs} & \textbf{\#params} & \textbf{\#train} & \textbf{\#test} & \textbf{time} & \textbf{top-1} & \textbf{top-5} \\ \hline
    English & 4,342 &      30.50 & 3,236,728 & 807,447 & 370 & 13.64\% & 26.63\% \\ %20160116-130559-309b [yael]
    Spanish & 2,382 &      20.84 & 1,085,678 & 270,400 & 106 & 13.86\% & 27.98\% \\ %20160125-234016-5f29 [yael]
    Italian & 1,561 & \,\,\,7.57 &   602,424 & 149,901 &  57 & 17.46\% & 32.35\% \\ %20160125-083124-80eb [yael]
    French  & 1,115 & \,\,\,7.12 &   462,522 & 115,112 &  36 & 16.76\% & 34.27\% \\ %20160519-200707-deff [gpu-pool/dig]
    German  &   275 & \,\,\,6.26 &   108,744 &  27,048 &  11 & 31.08\% & 54.10\% \\ %20160519-205133-7e79 [gpu-pool/dig2]
    Chinese &   243 & \,\,\,6.22 &   102,740 &  25,575 &  10 & 25.96\% & 47.11\% %20160520-080156-53b9 54604 [gpu-pool/dig2]
  \end{tabular}
  \caption{Classification accuracies of the new ANP detector banks (\%) using an Inception-based architecture\cite{szegedy_2015} with training walltimes (hours) and number of parameters (millions). The classes, training and testing sets all match those from Table \ref{tab:old_classifiers}, i.e.~those used in the original MVSO release\cite{jou_2015}.}
  \label{tab:new_classifiers}
\end{table}

Compared to the previous detector banks in Table \ref{tab:old_classifiers} which we fine-tuned from an affectively biased set of network weights \cite{chen_2014_dsb}, training takes noticeably longer in most cases since we now fine-tune from a model trained on ILSVRC.
Nonetheless, we observe that we are able to get improved ANP classification rates on most languages with Inception \cite{szegedy_2015}, particularly when there is more data for the networks to take advantage of; for example, with English ANP detection, we get a 35.05\% relative improvement at top-1.
Overall, except for French, we see a consistent improvement at top-5, indicating that more semantically relevant ANPs are being surfaced into the top ranks by Inception compared to the CaffeNet model.
While we do not see this same consistency at top-1, given that these detectors currently treat all ANPs as if they came from a flat taxonomy, many of the ANPs may in fact be semantically close to each other as indicated by the hierarchical grouping studied in the original ontology construction process \cite{jou_2015}.
As a result, from a purely empirical standpoint, a top-1 accuracy metric may treat a `pretty flower' prediction for a `beautiful flower'-labeled image to be a misclassification.
A top-5 metric in these ontology-structure-agnostic settings then may be a more reasonable indicator of detector bank performance than a top-1 performance metric.

\subsection{Tag-pool Multilingual ANP Detectors}
\label{ssec:tagpool}

\begin{figure}[t]
  \centering
  \begin{subfigure}[b]{0.49\textwidth}
    \centering
    \includegraphics[height=1.7in]{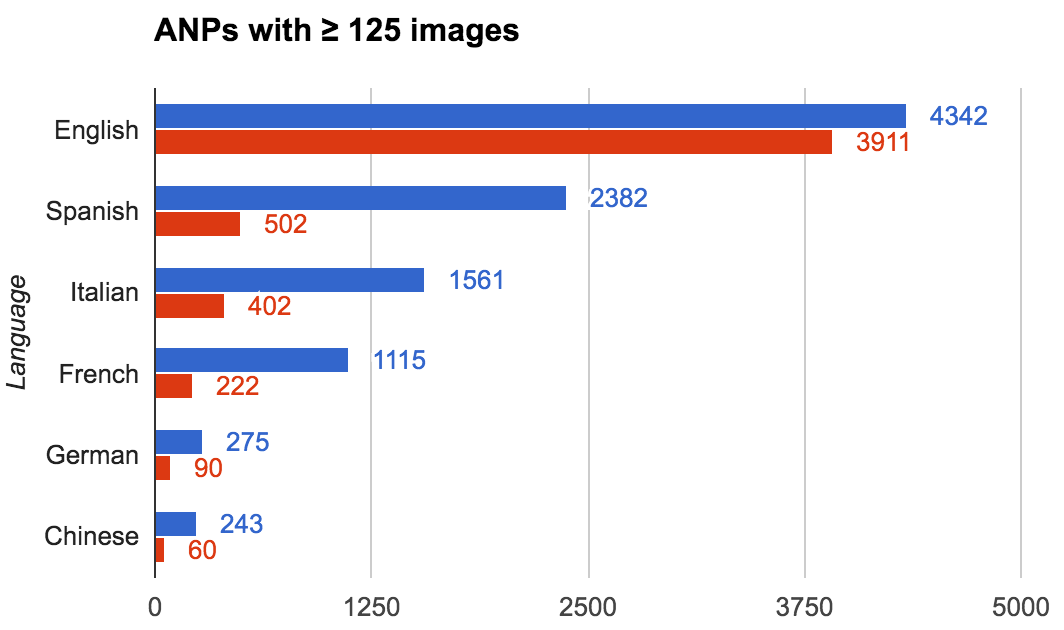}
    \caption{}
    \label{fig:mvso_tag_anps}
  \end{subfigure}
  \begin{subfigure}[b]{0.49\textwidth}
    \centering
    \includegraphics[height=1.7in]{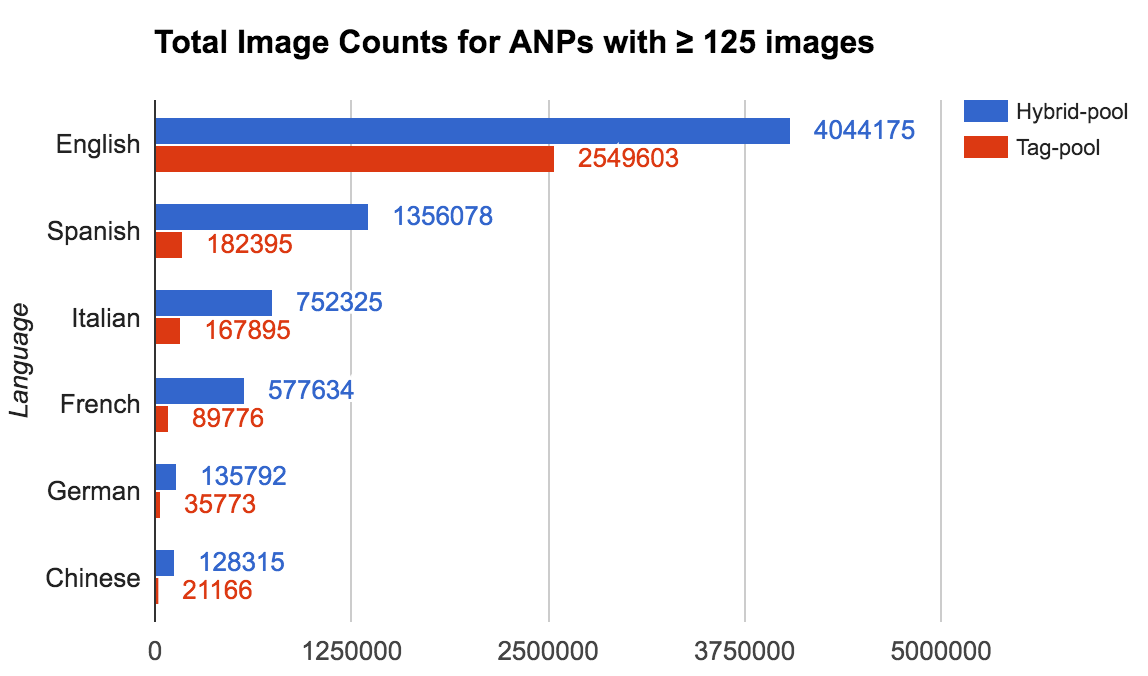}
    \caption{}
    \label{fig:mvso_tag_images}
  \end{subfigure}
  \caption{Hybrid \texttt{vs.}~Tag-pool MVSO Coverage. In \textbf{(a)}, we show the coverage of ANPs and in \textbf{(b)}, images, across the six major languages in MVSO\cite{jou_2015}. English suffers the least from the tag restriction while all other language experience a large drop in ANP and image coverage.}
  \label{fig:mvso_tag_anp_imgs}
\end{figure}

Since the image corpus in the hybrid-pool setting, where images come from both tag and free-text search on Flickr, can be noisy, we perform a separate set of experiments where we train our ANP detector banks just from the subset of images that come from tag-restricted queries on Flickr.
We call this the \emph{tag-pool} of MVSO images.
Though it is a subset, each ANP still has a maximum of 1,000 images to learn from and we split the dataset as before 80/20\% per ANP.
However, since we still enforce the requirement that there are at least 125 images per ANP, some languages experienced significant decrease in ANP coverage.
In Figure \ref{fig:mvso_tag_anp_imgs}, we show the change in ANP and image coverage when going to a tag-restricted subset in MVSO.
We observe that English experiences a small 9.92\% loss in ANPs and a 63.04\% loss in image count when restricting to tag-only queries while all other languages experience about 80\% coverage loss for both ANPs and images.
Though the ANPs in MVSO are useful given their pervasive occurrence in social media, this indicates that there are a fair number of images in hybrid-pool have a considerable amount of weak supervision.
At the same time, it is worth noting that it is wrong to assume all images from the free-text search are not useful for visual recognition of adjective-noun pairs.

In Table \ref{tab:tag_classifiers} below, we show the detector bank performances for the six major languages using the tag-restricted subset of images with CaffeNet-styled \cite{jia_2014} architectures.
To train these tag-pool ANP detectors, we follow training scheme from the hybrid-pool ANP detectors in the original MVSO release \cite{jou_2015} for 1-to-1 comparison, pre-training with DeepSentiBank \cite{chen_2014_dsb} using Caffe \cite{jia_2014} with the same optimization strategies, with the only slight difference being that we use a different GPU, i.e.~980 versus Titan X, and use the true training mean image instead of the ILSVRC mean.
Again, note that the number of ANP classes each detector learns over has reduced compared to those seen earlier in Table \ref{tab:old_classifiers} due to the tag-based pool restriction.

\begin{table}[h]
  \centering
  \begin{tabular}{lccccccc}
    \textbf{Language} & \textbf{\#ANPs} & \textbf{\#params} & \textbf{\#train} & \textbf{\#test} & \textbf{time} & \textbf{top-1} & \textbf{top-5} \\ \hline
    English & 3,911 & 72.89 & 2,294,411 & 255,192 &  85 & 19.00\% & 33.81\% \\
    Spanish &   502 & 58.92 &   164,119 &  18,276 &  12 & 29.07\% & 52.86\% \\
    Italian &   402 & 58.52 &   151,088 &  16,807 &   9 & 33.69\% & 55.76\% \\
    French  &   222 & 57.78 &    80,790 &   8,986 &   6 & 34.70\% & 63.32\% \\
    German  &    90 & 57.24 &    32,195 &   3,578 & 2.5 & 47.04\% & 74.62\% \\
    Chinese &    60 & 57.11 &    19,044 &   2,122 & 1.5 & 45.05\% & 71.35\%
  \end{tabular}
  \caption{Classification performances of the CaffeNet-based\cite{jia_2014} ANP detectors (\%) using images retrieved by restricting Flickr API queries to tag-based search. Note that walltimes (hours) are for running a fixed 80 epochs.}
  \label{tab:tag_classifiers}
\end{table}

\begin{table}[t]
  \centering
  \begin{tabular}{lc|ccc||ccc}
    \textbf{Language} & \textbf{\#ANPs} & \textbf{\#test} & \textbf{top-1} & \textbf{top-5} & \textbf{\#test} & \textbf{top-1} & \textbf{top-5} \\ \hline
    English & 3,911 & 756,243 & 10.47\% & 21.67\% & 51,062 & 12.04\% & 24.58\% \\
    Spanish &   502 &  91,002 & 14.86\% & 30.48\% &  3,740 & 17.91\% & 34.39\% \\
    Italian &   402 &  64,668 & 22.51\% & 39.44\% &  3,386 & 24.25\% & 42.85\% \\
    French  &   222 &  38,107 & 21.78\% & 43.10\% &  1,835 & 25.29\% & 46.54\% \\
    German  &    90 &  13,249 & 36.18\% & 58.99\% &    769 & 39.14\% & 62.68\% \\
    Chinese &    60 &   9,093 & 29.47\% & 48.78\% &    412 & 34.95\% & 51.94\%
  \end{tabular}
  \\
  \hspace{1.53in}(a)\hspace{1.6in}(b)
  \caption{Classification performances of the original CaffeNet-based \cite{jia_2014} ANP detectors (\%) from the ACM Multimedia 2015 release \cite{jou_2015}, i.e.~same detectors as in Table \ref{tab:old_classifiers}, but now evaluated only on the test images \textbf{(a)} of the tag-pool in Table \ref{tab:tag_classifiers}, and \textbf{(b)} intersection of both the hybrid-pool and tag-pool test sets.}
  \label{tab:hybrid_on_tag_eval}
\end{table}

At first glance, the top-1 and top-5 accuracy rates seem much higher than those we observed with a similar architecture in Table \ref{tab:old_classifiers}.
However, it is still difficult to determine if restricting to the tag-pool actually contributed to more reliable labels and thus better performing ANP detectors, especially since the number of output classes has changed.
As a result, we evaluated the hybrid-pool ANP detectors from \S\ref{ssec:hybridpool} on the test set of the tag-restricted pool of images as well as the the intersection of the test sets.
Since random shuffling was done during train/test splitting for both hybrid- and tag-pool image datasets, the latter ensures that no images in the test set (\#test) were used in either model's training.
The results from this evaluation are shown in Table \ref{tab:hybrid_on_tag_eval}.
We observe that in both cases, the classification performances are both lower than those in Table \ref{tab:tag_classifiers} by about 10\% absolute on each language, indicating that the tag-pool ANP detectors do indeed achieve a comparably higher accuracy on the same set of ANPs.
In addition, for an even closer comparison, we tested the trained tag-pool ANP detectors on the intersection of the hybrid-pool and test-pool test sets from Tables \ref{tab:old_classifiers} and \ref{tab:tag_classifiers}, respectively, as shown in Table \ref{tab:classifiers4}.
The results here are thus most comparable to those found in Table \ref{tab:hybrid_on_tag_eval}(b).
Although there are significantly less test images than with Table \ref{tab:old_classifiers}, \ref{tab:tag_classifiers}, or \ref{tab:hybrid_on_tag_eval}(a), we observe that the top-1 classification rates for the tag-pool English ANP detector bank is a relative 59.21\% improvement over the hybrid-pool English ANP detector bank on a comparable dataset.
ANP detection in all other language likewise see an improvement in the evaluation using the tag-pool ANP detectors. 

From the evaluations in Tables \ref{tab:tag_classifiers}-\ref{tab:classifiers4}, we believe we can conclude that training ANP detector banks using the tag-pool have led to significantly better ANP detectors.
The trade-off here, however, is that we have dramatically less images to train on and also a significantly reduced coverage of ANPs per language, except in the case of English.
Nonetheless, all this is serves as concrete evidence that tag-based labels are more reliable than those mined from the metadata overall since users are more likely to use tags to describe content that is visually present in the image.
While this does not mean that all labeled images gathered from free-text metadata context are wrong, it does reduce the trust that should be placed on their ability to represent elements visually present in the associated image.
We believe that in the future development of ANP detectors factoring such label reliability will allow for a greater coverage of ANPs without sacrificing in the area of data sparsity or the classification performance of ANP detector banks.

\begin{table}[h]
  \centering
  \begin{tabular}{lcccc}
    \textbf{Language} & \textbf{\#ANPs} & \textbf{\#test} & \textbf{top-1} & \textbf{top-5} \\ \hline
    English & 3,911 & 51,062 & 19.17\% & 34.01\% \\
    Spanish &   502 &  3,740 & 28.85\% & 52.75\% \\
    Italian &   402 &  3,386 & 32.75\% & 54.90\% \\
    French  &   222 &  1,835 & 35.80\% & 64.20\% \\
    German  &    90 &    769 & 49.41\% & 74.51\% \\
    Chinese &    60 &    412 & 41.50\% & 67.23\%
  \end{tabular}
  \caption{Classification performances of the CaffeNet-based\cite{jia_2014} ANP detectors (\%) using tag-restricted images but now evaluated only on the set of test images intersecting the hybrid pool of Table \ref{tab:old_classifiers}, i.e.~original MVSO data release\cite{jou_2015}, and tag-restricted pool of Table \ref{tab:tag_classifiers}. These tag-restricted performances are thus most comparable to the results reported in Table \ref{tab:hybrid_on_tag_eval}(b).}
  \label{tab:classifiers4}
\end{table}

\section{CONCLUSION}
\label{sec:conclusion}

In this technical report, we have presented new ANP detector banks for MVSO for six major languages using a more modern network architecture than in the original MVSO release.
These detectors have a reduced model memory footprint compared to their predecessors and achieve very competitive or even significantly better ANP detection performance, e.g.~a 35.05\% improvement for English ANP detection.
In addition, we presented ANP detectors that use a restricted subset of images and ANPs that originate only from tag-based Flickr API retrieval results, which resulted in much higher ANP classification performances due to improved label quality.
We also publicly release the new Inception-based ANP detector banks, tag-pool CaffeNet ANP detector banks, and the list of image URLs corresponding to the tag-restricted MVSO image subset used in this report.

In the future, we would like to investigate additional fine-tuning strategies for training ANP detectors.
For example, in the original release \cite{jou_2015}, we noted that fine-tuning from DeepSentiBank \cite{chen_2014_dsb} was done in the hopes that we initialized the models from an affectively biased source; however, it is not empirically clear whether such a biasing actually helps or hurts detector performance.
In addition, we would like to investigate and benchmark our ANP detectors against other popular network architectures.
%Further still, there are also promising directions to be pursued in pure adjective detection and joint target modeling \cite{narihira_2015,jou_2016_xres} of affective mid-level representations.
Also, we plan to investigate how we might integrate the semantic hierarchy in MVSO to train ANP detectors that exploit the tree-structure.

\acknowledgments % equivalent to \section*{ACKNOWLEDGMENTS}
We would like to thank Miriam Redi, Mercan Topkara, Nikolaos Pappas and Tao Chen from Multilingual Visual Sentiment Ontology (MVSO) team for their continued support and insightful discussions.
Also, we thank Margaret Yuying Qian for performing some of the early tag-restricted subset partitioning and statistics.

% References
\bibliography{mvso}

\begin{thebibliography}{10}

\bibitem{jou_2015}
Jou, B., Chen, T., Pappas, N., Redi, M., Topkara, M., and Chang, S.-F.,
  ``Visual affect around the world: {A} large-scale multilingual visual
  sentiment ontology,'' in [{\em ACM Multimedia
  (MM)}{\nolinebreak\hspace{0.1em}]},  (2015).

\bibitem{krizhevsky_2012}
Krizhevsky, A., Sutskever, I., and Hinton, G.~E., ``Image{N}et classification
  with deep convolutional neural networks,'' in [{\em Neural Information
  Processing Systems (NIPS)}{\nolinebreak\hspace{0.1em}]},  (2012).

\bibitem{chen_2014_dsb}
Chen, T., Borth, D., Darrell, T., and Chang, S.-F., ``Deep{S}enti{B}ank:
  {V}isual sentiment concept classification with deep convolutional neural
  networks,'' {\em arXiv preprint arXiv:1410.8586}  (2014).

\bibitem{szegedy_2015}
Szegedy, C., Liu, W., Jia, Y., Sermanet, P., Reed, S., Anguelov, D., Erhan, D.,
  Vanhoucke, V., and Rabinovich, A., ``Going deeper with convolutions,'' in
  [{\em Computer Vision and Pattern Recognition
  (CVPR)}{\nolinebreak\hspace{0.1em}]},  (2015).

\bibitem{picard_1997}
Picard, R.~W.,  [{\em Affective Computing}{\nolinebreak\hspace{0.1em}]}, MIT
  Press (1997).

\bibitem{borth_2013_vso}
Borth, D., Ji, R., Chen, T., Breuel, T., and Chang, S.-F., ``Large-scale visual
  sentiment ontology and detectors using adjective noun pairs,'' in [{\em ACM
  Multimedia (MM)}{\nolinebreak\hspace{0.1em}]},  (2013).

\bibitem{li_2016}
Liu, H., Jou, B., Chen, T., Topkara, M., Pappas, N., Redi, M., and Chang,
  S.-F., ``Complura: {E}xploring and leveraging a large-scale multilingual
  visual sentiment ontology,'' in [{\em Intl Conf. on Multimedia Retrieval
  (ICMR)}{\nolinebreak\hspace{0.1em}]},  (2016).

\bibitem{pappas_2016}
Pappas, N., Redi, M., Topkara, M., Jou, B., Liu, H., Chen, T., and Chang,
  S.-F., ``Multilingual visual sentiment concept matching,'' in [{\em Intl
  Conf. on Multimedia Retrieval (ICMR)}{\nolinebreak\hspace{0.1em}]},  (2016).

\bibitem{plutchik_1980}
Plutchik, R.,  [{\em Emotion: {A} Psychoevolutionary
  Synthesis}{\nolinebreak\hspace{0.1em}]}, Harper \& Row (1980).

\bibitem{jou_2016_senticart}
Jou, B., Qian, M.~Y., and Chang, S.-F., ``{SentiCart}: {C}artography \&
  geo-contextualization for multilingual visual sentiment,'' in [{\em ACM
  International Conference on Multimedia Retrieval
  (ICMR)}{\nolinebreak\hspace{0.1em}]},  (2016).

\bibitem{jia_2014}
Jia, Y., Shelhamer, E., Donahue, J., Karayev, S., Long, J., Girshick, R.,
  Guadarrama, S., and Darrell, T., ``Caffe: {C}onvolutional architecture for
  fast feature embedding,'' in [{\em ACM Multimedia
  (MM)}{\nolinebreak\hspace{0.1em}]},  (2014).

\bibitem{borth_2013_sb}
Borth, D., Chen, T., Ji, R., and Chang, S.-F., ``Senti{B}ank: {L}arge-scale
  ontology and classifiers for detecting sentiment and emotions in visual
  content,'' in [{\em ACM Multimedia (MM)}{\nolinebreak\hspace{0.1em}]},
  (2013).

\bibitem{imagenet}
Russakovsky, O., Deng, J., Su, H., Krause, J., Satheesh, S., Ma, S., Huang, Z.,
  Karpathy, A., Khosla, A., Bernstein, M., Berg, A.~C., and Fei-Fei, L.,
  ``{ImageNet Large Scale Visual Recognition Challenge},'' {\em Intl Journ. of
  Computer Vision (IJCV)}~{\bf 115}(3) (2015).

\bibitem{lin_2014}
Lin, M., Chen, Q., and Yan, S., ``Network in network,'' {\em arXiv preprint
  arXiv:1312.4400}  (2014).

\end{thebibliography}
\bibliographystyle{spiebib} % makes bibtex use spiebib.bst
% ACM needs 'a single self-contained file'!

%\balancecolumns % GM June 2007
\end{document}